\documentclass[11pt]{article}
\usepackage{moriond,epsfig}

\def\be{\begin{equation}}
\def\ee{\end{equation}}
\def\bea{\begin{eqnarray}}
\def\eea{\end{eqnarray}}

\begin{document}
\hfill hep-ph/0406180

\hfill IPPP-04-30

\hfill DCPT-04-60

\vspace*{3.2cm}
\title{SYNERGY OF COMBINED LHC AND LC ANALYSES IN SUSY SEARCHES}

\author{ G. MOORTGAT-PICK\footnote{email address: g.a.moortgat-pick@durham.ac.uk} }

\address{Institute for Particle Physics Phenomenology, University of Durham,
South Road,\\
Durham DH1 3LE, England}

\maketitle\abstracts{We present a case study for the
synergy of combined LHC and LC analyses in Susy searches where
simultaneous running of both machines is very important. 
In case that only 
light non-coloured Susy particles are accessible at a Linear
Collider with an initial energy of $\sqrt{s}=500$~GeV, the precise analysis 
at the LC nevertheless
leads to an
accurate Susy parameter determination. This allows the prediction of
heavy Susy particles. Providing these LC results as input for LHC analyses
could be crucial for the identification of signals
resulting in a direct measurement of the heavy neutralinos. 
These results provide an important consistency test of the 
underlying model. Furthermore, 
feeding back the LHC results into LC analyses leads to an
improvement in the
parameter determination.}

\section{Introduction}\label{intro}
\vspace{-.3cm}
One of the best motivated extensions of the Standard Model (SM) is
Supersymmetry (Susy). Therefore Susy searches will get a large
weight at the new physics searches at
the Large Hadron Collider (LHC), whose first run is
foreseen for 2007 and where data taking is expected to continue for about 20
years. Since Susy, if realised in nature, has to be a broken
symmetry, a large amount of new parameters enter in addition to the 19 SM
free parameters. They have to be precisely determined in order to
reveal the underlying structure of the model. In the
Minimal Supersymmetric Standard Model (MSSM) one is faced with
around 105 new free parameters. Therefore care is required to 
impose as little model assumptions
on the experimental analyses as possible. 
Due to the clear signatures at the Linear Collider (LC) 
a largely model-independent determination of masses,
couplings, mixing angles, phases and quantum numbers 
can be done in the general MSSM parameter space.  Therefore significant help
for LHC analyses via particle mass measurements and 
predictions from analyses at a LC 
is expected.
Searches for light new particles at a LC with
a first energy phase of
$\sqrt{s}=500$~GeV, which could start running at 2015, may be crucial.
The mass predictions from the LC may lead 
to a precise mass
measurement of heavy new particles via the LHC analyses.
Therefore, the interplay of both colliders will
provide a powerful
consistency check of the model at an early stage of both experiments
and may outline future strategies for new physics searches at the LHC. 
Furthermore, feeding back the LHC results 
increases the accuracy 
of the fairly model-independent Susy parameter
determination at the LC~\cite{DKMNP}.

Interest in working out such examples of a synergy between
both experiments was initiated by the world-wide LHC/LC study
group \footnote{webpage: http://www.durham.ac.uk/ $\tilde{}$ georg/lhclc},
founded in 2002. The results so far are summarised in a working group report
\cite{lhclc}.

We choose as a representative example the Susy reference scenario
SPS1a \cite{SPS} which is a quite favourable parameter point for both machines
where already some experimental simulations exist. While SPS1a is based on
an mSugra scenario, i.e.\ the 
Susy breaking is transmitted via gravitational interactions, 
for the further procedure we do not
make any assumptions depending on the Susy breaking scenario.
In the following we mainly concentrate on the non-coloured particle sector.

\section{Susy studies at the LHC}\label{chap1}
\begin{figure}
\hskip 0.5cm
\psfig{figure=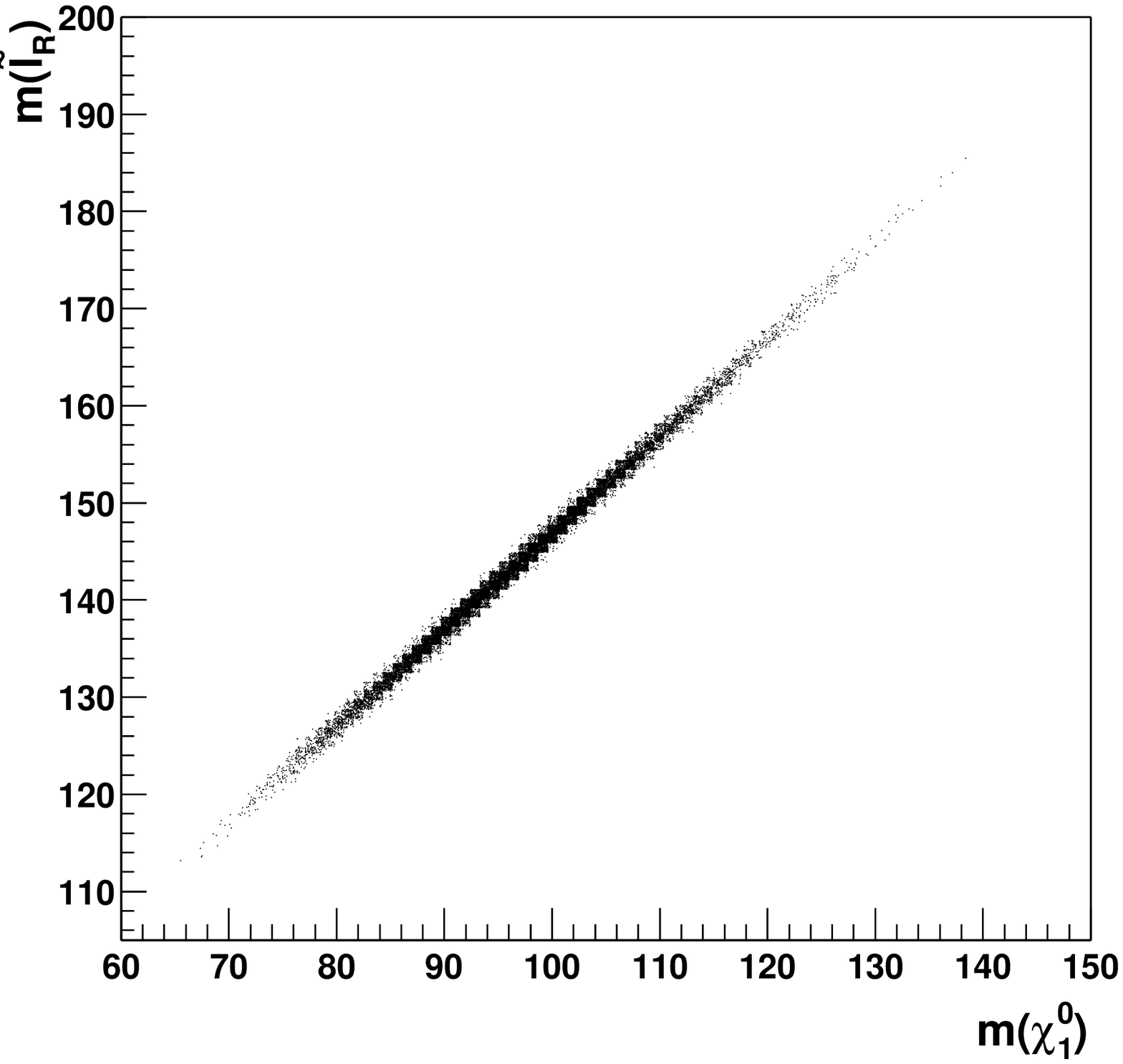,height=2.5in,width=3in}
\hskip .5cm
\psfig{figure=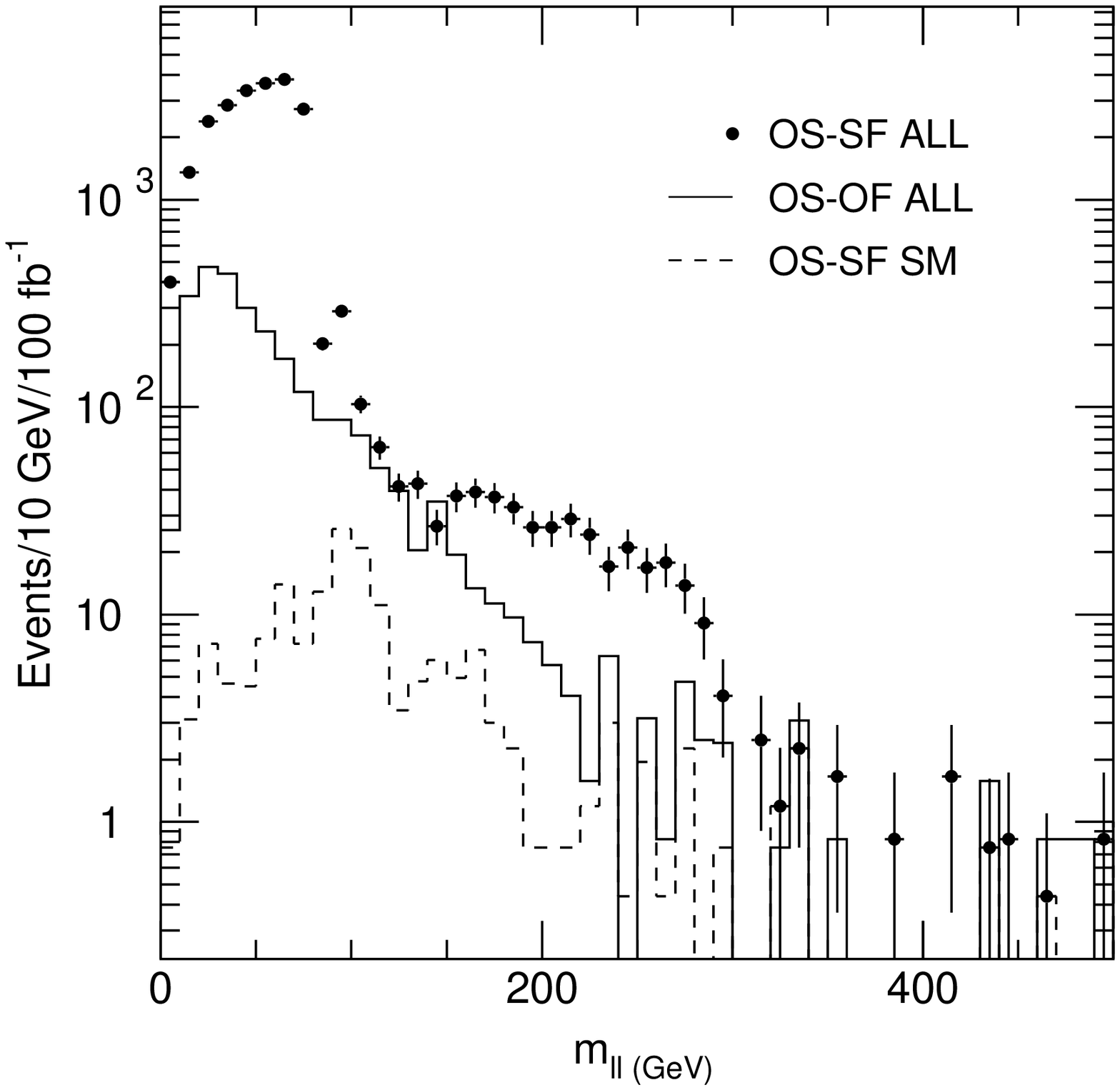,height=2.5in,width=3in}
\hskip 0.5cm
\vskip -1.3cm
\caption{Mass measurements at the LHC: sensitivity of $m_{\tilde{\ell}_R}$ to the mass of the 
LSP $m_{\tilde{\chi}^0_1}$ in the reference scenario SPS1a$^{2,4}$ (left) and
invariant mass spectrum of the heavy neutralino/chargino decay chains$^5$ (right).
The dilepton OS-SF lepton edge of $\tilde{\chi}^0_4$ is the edge between
200~GeV$<m_{ll}<$400~GeV.
\label{fig-lhc}
}
\end{figure}

\vspace{-.3cm}
Detailed simulations of the LHC capabilities for the reference point SPS1a
were  carried out~\cite{LHCstudy};
the masses of the Susy particles
can in most cases 
only be studied by analysing complicated decay chains, like
\begin{equation}
\tilde{q}_L\to \tilde{\chi}^0_2 q \to \tilde{\ell}^{\mp}_R \ell^{\pm} q\to \tilde{\chi}^0_1
\ell^{\mp} \ell^{\pm} q,
\label{eq_chain}
\end{equation}
which might be difficult to resolve.  The precise reconstruction of the states in
the decay chains requires in particular the knowledge of the mass of the
lightest Susy particle (LSP), which is often assumed to be stable. As an example for the 
strong sensitivity to  $m_{\tilde{\chi}^0_1}$ we
show in  Fig.~\ref{fig-lhc} (left) the determination of $m_{\tilde{\ell}_R}$
 in dependence of $m_{\tilde{\chi}^0_1}$ \cite{LHCstudy}.
Applying a
joint fit of various kinematic 'edges' yields an overconstraint
system and leads to an indirect knowledge on $m_{\tilde{\chi}^0_1}$. 
However, some assumptions about particle identities have to be made. 
Using LC results 
leads to an higher accuracy in determining the masses and provides model-independent 
consistency tests.

In our reference point simulations were done to determine also
the gaugino/higgsino particles.
The second lightest neutralino can be identified in the
opposite sign-same flavour signal (OS-SF) with an uncertainty of about
$\delta m_{\tilde{\chi}^0_2}=4.7$~GeV. The main background from $\tilde{\chi}^{\pm}_1$
decays yields an equal number of (OS-SF) and Opposite-Sign Opposite-Flavour (OS-OF)
leptons pairs and can thus be separated by subtraction
 \cite{Giacomo}.

The heavy charged and the neutral gaugino/higgsino particles are
nearly mass degenerate.  The resolution of the corresponding edges is therefore
particularly difficult. The neutralino $\tilde{\chi}^0_3$ is nearly a
pure higgsino and does not couple to squarks, therefore only
$\tilde{\chi}^{\pm}_2$ and $\tilde{\chi}^0_4$ are accessible. The
competing decay chains in this case are
\begin{eqnarray}
&&\tilde{\chi}^0_4 q \to \tilde{\ell}^{\pm}_R \ell^{\mp} q\to
\tilde{\chi}^0_1 \ell^{\pm}  \ell^{\mp} q
\label{eq_chain1}\\
&&\tilde{\chi}^0_4 q \to \tilde{\ell}^{\pm}_L \ell^{\mp} q\to
\tilde{\chi}^0_1 \ell^{\pm} \ell^{\mp} q \mbox{\quad or \quad}\tilde{\chi}^0_2
\ell^{\pm} \ell^{\mp} q\label{eq_chain2}\\ 
&&\tilde{\chi}^{\pm}_2 q' \to
\tilde{\nu}_{\ell} \ell^{\pm} q'\to \tilde{\chi}^{\pm}_1 \ell^{\mp}  \ell^{\pm} q'
\label{eq_chain3}
\end{eqnarray}
In combination with measured invariant masses one can derive the OS-SF
signal of the heavy particle with $\delta(m)=5.1$~GeV, and under specific assumptions one
can interpret the edge as that of the 
$\tilde{\chi}^0_4$ particle \cite{LHCstudy,Giacomo}.
\section{Susy studies at the LC}
\vspace{-.3cm}
Precise simulations for the mass measurements of the sleptons and the
light charginos and neutralinos at the Linear Collider have also been
done for the parameter point SPS1a \cite{Martyn,Ball}, the results
are given Table~\ref{tab_mass_LC}.  Particularly interesting is the
high accuracy in the determination of $m_{\tilde{\chi}^0_1}$ with
$\delta(m_{\tilde{\chi}^0_1})=0.05$~GeV from $\tilde{e}_R$ decays, but
also the accuracy $\delta (m_{\tilde{\chi}^{\pm}_1})=0.55$~GeV and
$\delta (m_{\tilde{\chi}^{0}_2})=1.2$~GeV are important. Due to $\tan\beta=10$ 
in the chosen parameter point the
light chargino $\tilde{\chi}^{\pm}_1$ as well as the second lightest
neutralino $\tilde{\chi}^0_2$ decay both mainly into $\tilde{\tau}$'s
producing a signal similar to that of stau-pair production.  The
final states from $\tilde{\chi}^+_1\tilde{\chi}^-_1$ and
$\tilde{\chi}^0_1\tilde{\chi}^0_2$ decays are the same (2 $\tau$+
missing energy), however with different topology. This feature allows
to separate the process to some extent exploiting e.g. suitable cuts
on the opening angle between the leptons.

The precise measurement of the Susy particle masses 
as well as the different cross sections
of this light particle spectrum alone leads to a precise determination of the fundamental
Susy parameters which govern the chargino-neutralino sector. Within the general MSSM,
from these parameters the masses of the heavier neutralinos and the heavier chargino can be 
predicted without further model assumptions.
\subsection{Strategy for Susy parameter determination}
\vspace{-.2cm}
We follow, as an example, 
a method described in \cite{ckmz} and take into account
in addition the simulated errors in the mass and cross section measurements \cite{DKMNP}.
The  mass matrix ${\cal M}_C$
of the charged gauginos $\tilde{W}^\pm$ and higgsinos $\tilde{H}^\pm$
depends on $M_2$, $\mu$, $\tan\beta$. The mass eigenstates
are the two charginos $\tilde{\chi}^\pm_{1,2}$. For real ${\cal M}_C$
the two unitary diagonalisation
matrices can be parameterised with two mixing angles
$\Phi_{L,R}$. The mass eigenvalues $m^2_{\tilde{\chi}^\pm_{1,2}}$ and the mixing
angles are given by the Susy parameters, see e.g.\cite{Choi}.
The cross section
$\sigma^\pm\{ij\}=\sigma(e^+e^-\to\tilde{\chi}^{\pm}_i
\tilde{\chi}^{\mp}_j)$ can be expressed as a function of
$(\cos 2 \Phi_{L,R},m^2_{\tilde{\chi}^\pm_{i}})$; the coefficients for
$\sigma^\pm\{11\}$ are explicitly given in Ref.~\cite{DKMNP}.
The chargino  cross sections are measured at
$\sqrt{s}=400$~GeV and 500 GeV with polarised beams
so that the mixing angles $\cos 2 \Phi_{L,R}$ can unambiguously
be determined. Together with $m_{\tilde{\chi}^{\pm}_1}$ the parameters $M_2$, $\mu$,
$\tan\beta$ can be derived.

The neutralino mixing matrix ${\cal M}_N$ depends on $M_1$, $M_2$, $\mu$ and
$\tan\beta$. Analytic expressions for the mass eigenvalues
$m^2_{\tilde{\chi}^0_{1,\ldots,4}}$ and the eigenvectors are e.g. given in
\cite{ckmz}. The characteristic equation of the mass matrix squared,
${\cal M}_N {\cal M}^{\dagger}_N$, can be written explicitly\cite{DKMNP}
as a quadratic equation of the U(1) gaugino mass parameter $M_1$.
Together with the kinematically
accessible cross sections for
the light neutralino production, $\sigma^0_{L,R}\{12\}$ and $\sigma^0_{L,R}\{22\}$,
a precise and unambiguous determination of $M_1$, $M_2$, $\mu$ and $\tan\beta$ can be performed
without assuming a specific Susy breaking scheme.
\subsection{Results}
\vspace{-.2cm}
We took into account the following uncertainties:\\
$\bullet$
The uncertainties in the mass measurement, see Table~\ref{tab_mass_LC}.\\
$\bullet$ With $\int {\cal L}=500$~fb$^{-1}$
at the LC,  we assume 100~fb$^{-1}$ per each polarisation configuration
and\\
\phantom{$\bullet$ }we take into account 1$\sigma$ statistical errors for the cross sections.\\
$\bullet$ The beam polarisation measurement is assumed
with an uncertainty
of $\Delta P(e^{\pm})/P(e^ {\pm})=0.5\%$.\\
$\bullet$ Since the chargino (neutralino) production is sensitive to
$m_{\tilde{\nu}_e}$ ($m_{\tilde{e}_{L,R}}$),
we include the experi-\\
\phantom{$\bullet$ }mental errors of their mass determination of
0.7~GeV (0.2~GeV, 0.05~GeV), see Table~\ref{tab_mass_LC}.\\
$\bullet$ Concerning the neutralino cross sections we
estimate the statistical error based on an experi-\\
\phantom{$\bullet$ }mental
simulation\cite{Ball}
yielding an efficiency of 25\% and include
an additional systematic error\\
\phantom{$\bullet$ }($\delta\sigma_{\mbox{bg}}$) which takes
into account the uncertainty in the background subtraction,
for details see~\cite{DKMNP}.\\

The dominant error in the cross sections
is the statistical error and reaches up to
$\sim 4\%$ for left-handed polarised beams  
and up to $\sim 16\%$ 
for right-handed polarised beams due to partially low rates, 
see Table~\ref{tab_cross_LC}. The other dominant error
is due to the uncertainty in the mass measurement of 
$\delta(m_{\tilde{\chi}^{\pm}_1})$ which results in an error of 
about 2-3\% in the cross sections. The errors caused by the uncertainty in the beam polarisation
$\Delta P(e^{\pm})/P(e^ {\pm})=0.5\%$ lead to errors $\ll 1\%$ for left-handed beams and up to
$\le 2\%$ for right-handed beams. Errors caused by the mass uncertainties of the exchanged particles
are $\ll 1\%$.

Such a precise analysis of the light particle spectrum
 leads to a very accurate determination of the underlying
fundamental Susy parameters.
We use a 3-parameter $\Delta \chi^2  =\sum_i |(O_i -\bar O_i)/\delta O_i|^2$ test,
yielding
\begin{equation}
M_1=99.1\pm 0.2,\quad M_2=192.7\pm 0.6,\quad \mu=352.8\pm 8.9,\quad \tan\beta=10.3\pm 1.5.
\label{eq_par_lc}
\end{equation}
Since the light particles in our reference scenario are mainly gaugino-like we
derive precise values for the 
gaugino mass parameters $M_{1,2}$, but less accurate values for the
higgsino mass parameter $\mu$ and
for $\tan\beta$. However, the determination of the parameters is sufficient to
predict the heavier chargino and neutralino masses with high precision:
\begin{equation}
m_{\tilde{\chi}^{\pm}_2}=378.8\pm 7.8,\quad m_{\tilde{\chi}^0_3}=359.2\pm 8.6,\quad
m_{\tilde{\chi}^0_4}=378.2\pm 8.1.
\label{eq_mass_pred}
\end{equation}

\section{Susy studies in combined LHC/LC analyses}
\vspace{-.3cm}
Feeding some results of the LC analysis, i.e.~the mass
predictions as well as the precisely measured masses of the light Susy
particles, $m_{\tilde{\chi}^0_1}$, $m_{\tilde{\chi}^{\pm}_1}$,
$m_{\tilde{e}_{L,R}}$, $m_{\tilde{\nu}}$, as input into the LHC
analysis leads to the following improvements at the LHC analysis:\\
$\bullet$ increase of statistical sensitivity due to the mass
predictions ('look elsewhere effect'), which\\ \phantom{$\bullet$
}could be crucial for the search for statistically marginal
signals;\\ $\bullet$ clear identification of the dilepton edge from
the $\tilde{\chi}^0_4$ decay chain;\\ $\bullet$ accurate measurement
of $m_{\tilde{\chi}^0_4}=377.87\pm 2.23$~GeV;\\ $\bullet$ the precise
identification of a dilepton edge right at the predicted mass with the
help of the\\ \phantom{$\bullet$ }LC means an important check of the
underlying Susy model at an early stage of both\\ \phantom{$\bullet$
}experiments, the LHC in combination with a LC$_{500}$;\\ $\bullet$
better accuracy also for $\delta(m_{\tilde{\chi}^0_2})=0.08$~GeV due
the precise knowledge on the LSP mass, $m_{\tilde{\chi}^0_1}$.

Using these improved results from the LHC analysis as input for further
analyses at the LC leads also to an improvement in the Susy parameter
determination. Since for our reference point the heavier neutralino states
are mainly higgsino-like, we increase thus in particular the accuracy on
the higgsino mass parameter, $\mu=352.4\pm 2.1$~GeV,  and
on $\tan\beta=10.2\pm 0.6$, see also Table~\ref{tab_par_com}.

\section{Conclusions}
\vspace{-.3cm}
Future experiments will face the task to unravel possible kinds
of physics beyond the SM. We have shown a representative case study where
searches for new physics models -- we have chosen Supersymmetry -- may
greatly benefit from the synergy of the combined analysis at the LHC
and at the LC in its first energy stage of $\sqrt{s}=500$~GeV.

We studied the prospects for resolving the Susy gaugino/higgsino sector.
We focused on the situation where only the light states
($\tilde{\chi}^0_1$, $\tilde{\chi}^0_2$, $\tilde{\chi}^\pm_1$) are
accessible at the first stage of the LC.  For a representative example of the
MSSM, we perform
a precise determination of the fundamental SUSY
parameters at the LC.
The masses of heavier chargino and neutralinos
can be subsequently predicted at the level of a few percent.

Concerning studies at the LHC 
the mass predictions from the LC analysis
lead to an increase of statistical sensitivity.
Together with a precise knowledge on the LSP mass and the light
slepton masses measured at the LC, the mass predictions
lead to a clear identification of the heavy
neutralinos in the corresponding decay chains at the LHC analysis, followed by
precise mass measurements of these heavy particles.

Measuring the heavy particles right at the predicted masses provides
an important check of the underlying Susy model.
Furthermore, feeding back the LHC results, i.e.\ the now
clearly identified and measured heavy electroweak particles,
into further analysis at the LC$_{500}$ leads to an even more accurate 
determination of the Susy parameters $M_1$, $M_2$, $\mu$ 
with an accuracy at the $\le O(1\%)$ level, and
an error on $\tan\beta$ of the order of $\le 10\%$. At this stage of accuracy radiative 
corrections become relevant in the electroweak sector \cite{Blank}, which will have to be taken into
account in future fits \cite{Fittino}.

The analysis has been performed within the general frame
of the unconstrained MSSM.  Our strategy does not rely on any
particular relations among the fundamental parameters, like the GUT or
mSUGRA relations, and therefore is applicable for arbitrary MSSM
parameters which lead to a phenomenology similar to the one studied.


\begin{table}[t]
\caption{Chargino, neutralino and slepton  masses in SPS1a, and
the simulated experimental errors at the LC$^{6,7}$.
It is assumed that the heavy
chargino and neutralinos are not observed at the first phase of the
LC operating at $\sqrt{s}\le 500$~GeV.
[All quantities are in GeV.]
\label{tab_mass_LC}}
\vspace{0.4cm}
\begin{center}
\begin{tabular}{|c|cc|cccc|ccc|}
\hline
& ${\tilde{\chi}^{\pm}_1}$ & ${\tilde{\chi}^{\pm}_2}$ &
${\tilde{\chi}^0_1}$ & ${\tilde{\chi}^0_2}$ &${\tilde{\chi}^0_3}$ &
${\tilde{\chi}^0_4}$ & ${\tilde{e}_R}$ &  ${\tilde{e}_L}$ &
${\tilde{\nu}_e}$ \\
\hline
mass &
176.03 & 378.50 & 96.17 & 176.59 & 358.81 & 377.87 & 143.0 & 202.1 & 186.0 \\
error &
0.55   &        & 0.05  & 1.2    &        &        & 0.05 & 0.2   & 0.7   \\
\hline
\end{tabular}
\end{center}
\end{table}

\begin{table}[t]
\caption{Susy parameters with 1 $\sigma$ errors derived from
the LC data collected at the first phase of operation and of
the combined analysis of the
LHC and LC$_{500}$ data with $\delta(m_{\tilde{\chi}^0_2})=0.08$~GeV and
$\delta(m_{\tilde{\chi}^0_4})=2.23$~GeV
derived from the LHC when using the LC input of $\delta(m_{\tilde{\chi}^0_1})=0.05$~GeV
${}^1$.
\label{tab_par_com}}
\vspace{0.4cm}
\begin{center}
\begin{tabular}{|c|cccc|} \hline
& $M_1$ & $M_2$ & $\mu$ & $\tan\beta$\\
\hline
\multicolumn{1}{|c|}{theo} & 99.1 & 192.7 & 352.4 & 10\\
LC$_{500}$& $99.1\pm 0.2$& $192.7\pm 0.6$& $352.8\pm 8.9$ &
$10.3\pm 1.5$\\
LHC+LC$_{500}$ & $99.1\pm 0.1$ & $192.7\pm 0.3$ &
{$352.4\pm 2.1$} & {$10.2\pm 0.6$} \\ \hline
\end{tabular}
\end{center}
\end{table}

\begin{table}[t]
\caption{Cross  sections of the processes $e^+e^-\to \tilde{\chi}^+_1 \tilde{\chi}^-_1$,
$e^+e^-\to \tilde{\chi}^0_1\tilde{\chi}^0_2$, $e^+e^-\to \tilde{\chi}^0_2\tilde{\chi}^0_2$
at $\sqrt{s}=400$~GeV and $\sqrt{s}=500$~GeV, respectively, with $P_{e^-}=-80\%$, $P_{e^+}=+60\%$
($\sigma_L$) and $P_{e^-}=+80\%$, $P_{e^+}=-60\%$ ($\sigma_R$) ${}^{1}$.
\label{tab_cross_LC}}
\vspace{0.4cm}
\begin{center}
\begin{tabular}{|c|cc|cc|}
\hline
&\multicolumn{2}{|c|}{$\sqrt{s}=400$~GeV}&\multicolumn{2}{c|}{$\sqrt{s}=500$~GeV}\\
\hline\hline
$\tilde{\chi}^{\pm}_1\tilde{\chi}^{\mp}_1$ & $\sigma_L=215$~fb &
$\sigma_R=6$~fb & $\sigma_L=504$~fb&$\sigma_R=15$~fb\\ \hline
$\tilde{\chi}^0_1\tilde{\chi}^0_2$ & $\sigma_L=148$~fb &
$\sigma_R=20$~fb & $\sigma_L=168$~fb&$\sigma_R=21$~fb\\ \hline
$\tilde{\chi}^0_2\tilde{\chi}^0_2$ & $\sigma_L=86$~fb &
$\sigma_R=2$~fb & $\sigma_L=217$~fb&$\sigma_R=6$~fb\\ \hline
\end{tabular}
\end{center}
\end{table}


\begin{figure}[t]
\psfig{figure=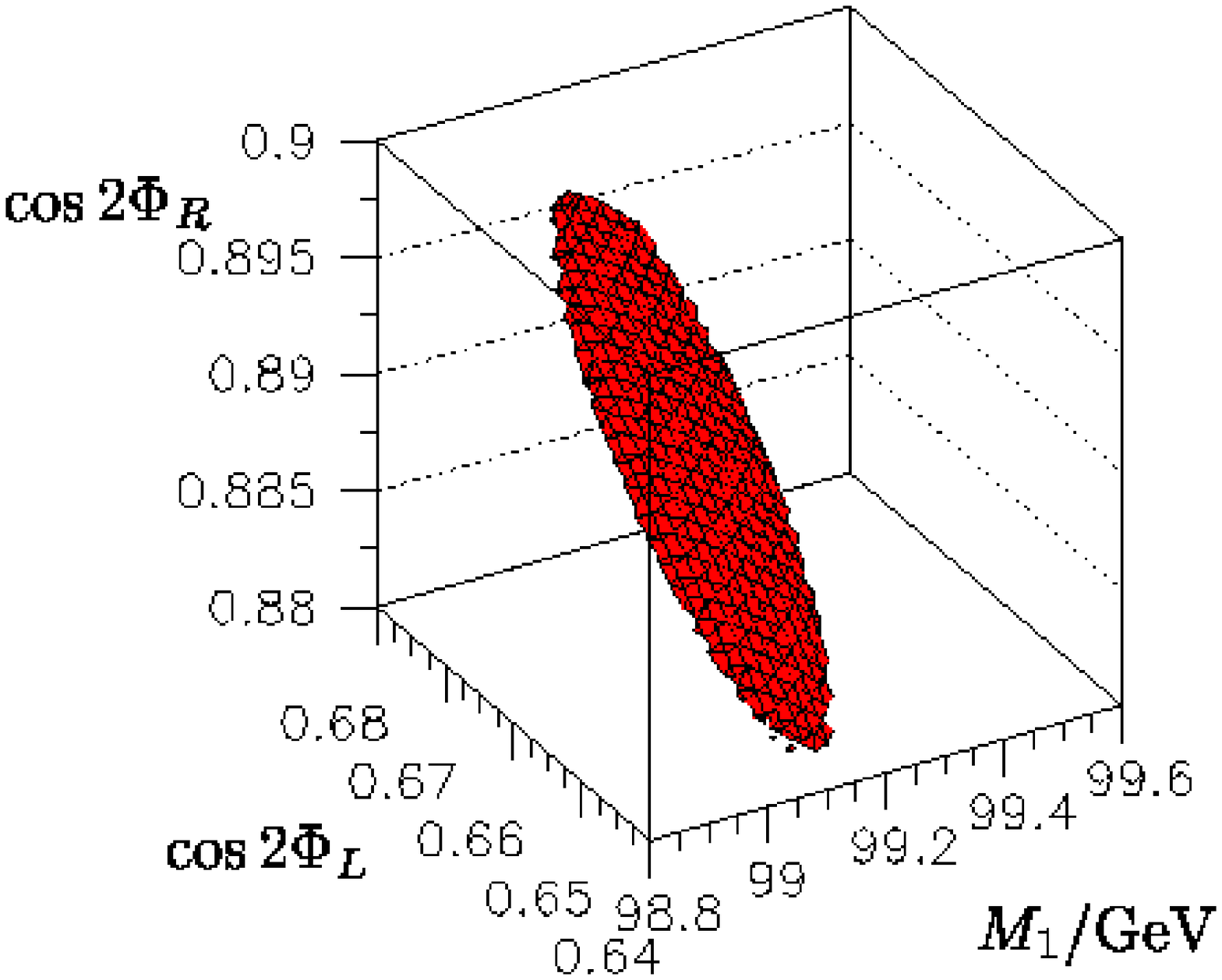,height=2.in,width=3in}
\hskip .5cm
\psfig{figure=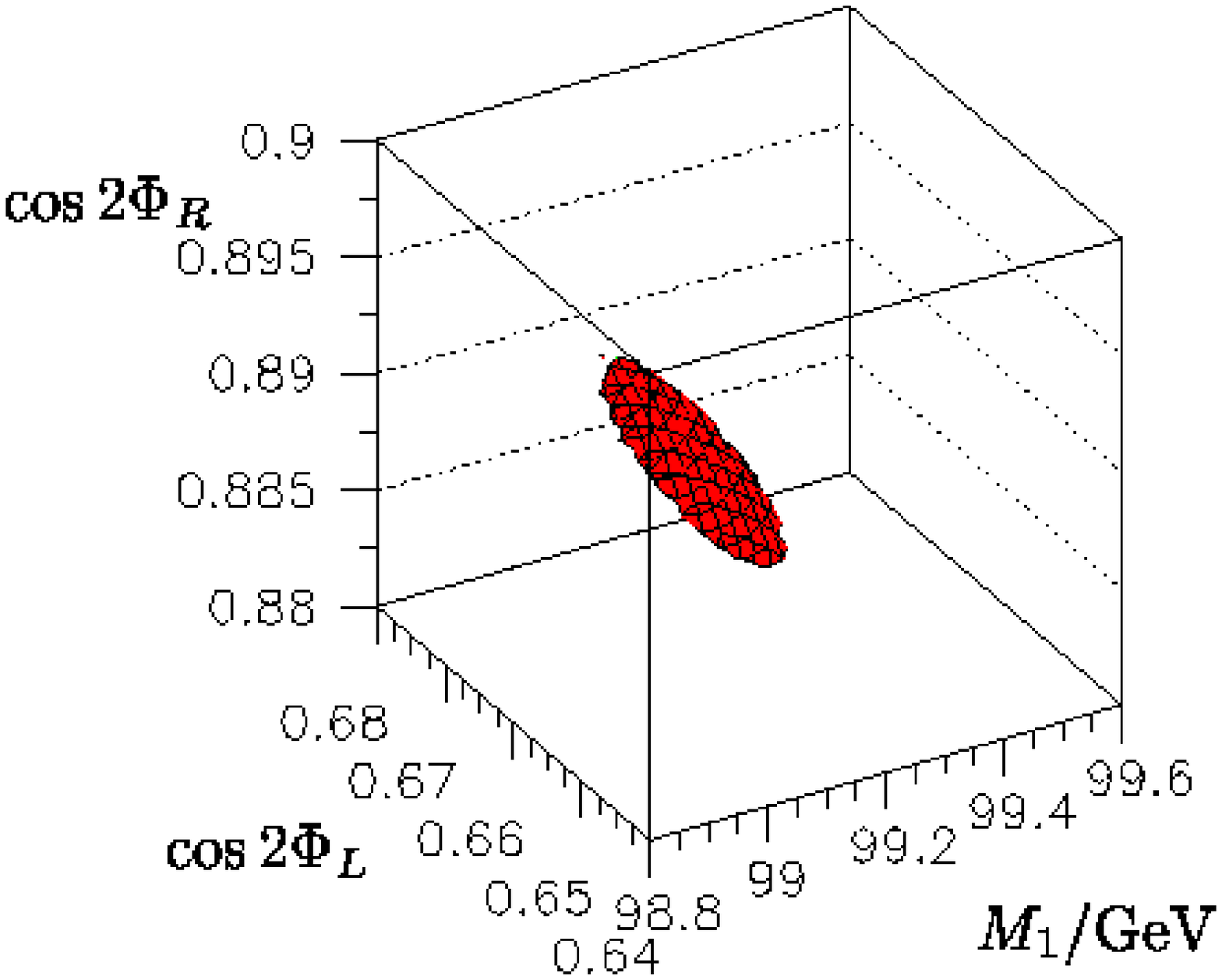,height=2.in,width=3in}
\caption{The $\Delta \chi^2=1$ contour in the
$\{M_1, \cos 2 \Phi_L, \cos 2 \Phi_R\}$ parameter space derived a) from the LC data and b) from
the joined analysis of the LC data and LHC data$^1$.
\label{fig-lc}}
\end{figure}

\section*{Acknowledgements}
\vspace{-.3cm}
The author would like to thank the organisers of the Moriond meeting, in particular
J.-M. Frere, for the invitation and an excellent meeting where the physics discussions
as well as the non-physics activities were very enjoyable. The author is very grateful to K. Desch, 
J. Kalinowski, M. Nojiri and G. Polesello for
the fruitful collaboration on the results presented here.
\section*{References}

\end{document}